\documentclass[aps,superscriptaddress,showpacs,nofootinbib,eqsecnum]{revtex4}

\usepackage{graphicx}
\usepackage{amsmath}
\usepackage{amsfonts}
\usepackage{graphicx}
\usepackage{epstopdf}
\usepackage{latexsym}
\usepackage{amssymb}
\usepackage{color}

\usepackage[center]{subfigure}

\begin{document}

 \newcommand{\bq}{\begin{equation}}
 \newcommand{\eq}{\end{equation}}
 \newcommand{\bqn}{\begin{eqnarray}}
 \newcommand{\eqn}{\end{eqnarray}}
 \newcommand{\nb}{\nonumber}
 \newcommand{\lb}{\label}
\newcommand{\PRL}{Phys. Rev. Lett.}
\newcommand{\PL}{Phys. Lett.}
\newcommand{\PR}{Phys. Rev.}
\newcommand{\CQG}{Class. Quantum Grav.}

\title{New Electrically Charged Black Hole in Higher Derivative Gravity }

\author{Kai Lin}
\email{lk314159@hotmail.com}
\affiliation{Instituto de F\'isica e Qu\'imica, Universidade Federal de Itajub\'a, MG, Brasil}
\affiliation{Instituto de F\'isica, Universidade de S\~ao Paulo, CP 66318, 05315-970, S\~ao Paulo, Brasil}
\affiliation{College of Physics and Electric Information, China West Normal University, Nanchong, Sichuan 637002, China}
\author{A. B. Pavan}
\email{alan@unifei.edu.br}
\affiliation{Instituto de F\'isica e Qu\'imica, Universidade Federal de Itajub\'a, MG, Brasil}
\author{G. Flores-Hidalgo}
\email{gfloreshidalgo@unifei.edu.br}
\affiliation{Instituto de F\'isica e Qu\'imica, Universidade Federal de Itajub\'a, MG, Brasil}
\author{E. Abdalla}
\email{eabdalla@usp.br}
\affiliation{Instituto de F\'isica, Universidade de S\~ao Paulo, CP 66318, 05315-970, S\~ao Paulo, Brasil}

\date{\today}

\begin{abstract}

In this paper, new electrically charged asymptotically flat black hole solutions are numerically derived in the context of higher derivative gravity. These solutions can be interpreted as generalizations of two different classes of non-charged asymptotically flat spacetimes: Schwarzschild and non-Schwarzschild solutions. Extreme black hole solutions and black holes with negative mass were found.
\end{abstract}


\maketitle

\section{Introduction}
\label{intro}

It is well know that quantized Einstein General Relativity Theory of gravity is nonrenormalizable. On the other hand, a way to remove the ultraviolet divergences in any quantum field theory is to modify its propagator by adding higher derivatives terms to the lagrangian. Following this idea, some years ago, Stelle  proposed to add all possible quadratic curvature invariants to the usual Einstein-Hilbert action \cite{Stelle} and he obtained a theory of quantum gravity free of ultraviolet divergences but as early recognized \cite{Pais}, the price of adding higher derivative terms is the introduction of unphysical ghost-like quanta in the spectrum of the theory.

Although unphysical ghosts, in general, violate unitarity and thus the probabilistic interpretation of quantum theory, there are some arguments that reinforce the idea that this is not a severe problem, as showed in \cite{Smilga}. In this case there are theories with ghosts but that preserve unitarity \cite{Smilga1}.  The key point in the approach of these works have been the study of the classical theory and whenever it is stable it is guaranteed the unitarity of the corresponding quantum states. Since black holes emerges as solutions of the classical Einstein field equations therefore it is expected that such objects could be of central role in establishing the unitarity of higher derivative modifications of Einstein General Relativity.

Motivated by above issues and also because black holes are important objects on their own, in this work we consider the search for electric charged black hole solutions in higher derivative modified gravity with additional curvature terms. Specifically, we consider additional Weyl and squared Ricci scalars as in a recent work \cite{LPPS}, where the authors obtained  numerically non-Schwarzschild static black hole solutions. However as far as we know, no electrically charged black hole have yet considered in this model.
Thus, searching for this kind of solution we have found new charged black hole solutions which, surprisingly, can not be reduced to Reissner-Nordstr\"om solution in the limit when the high derivative terms are less relevant.

This work is presented as follows. In section \ref{solutions}, we introduce the Maxwell field into the action of Einstein gravity with additional quadratic curvature terms and derive its equations of motion. After, the weak field limit solution is obtained and its behaviour is discussed. Later we derive new electrically charged asymptotically flat black hole solutions numerically and analyse their proprieties. Finally, in section \ref{conclusions} are presented our conclusions.

\section{Charged black hole in Higher Derivative gravity}
\label{solutions}

As it is known one of the worst difficulties in quantum gravity is the fact that quantized general relativity is nonrenormalizable. However, the present experiments support Einstein's gravitational theory and it means that a quantum gravity theory should satisfy all present experiments but correct the General Relativity at Planck scale. One ofthe simplest idea of constructing a renormalizable theory of gravity is by introducing higher derivative curvature terms into the gravitational action. In this way, many years ago,  Stelle has proved that Einstein-Hilbert action added with all possible quadratic curvature invariants is a renormalizable theory, though ghostlike modes are introduced on it \cite{Stelle}.

More recently, L\"u \emph{et. al} have considered a theory of gravity with quadratic Weyl and Ricci curvature invariants  \cite{LPPS}.
In this context they have found, numerically, a new spherically symmetric vacuum solution named non-Schwarzschild black hole and which admits positive and negative values for the the black hole mass. Thus, the next step would be to generalize this previous solution looking for a new electrically charged black hole solution.

In order to find a charged black hole we introduce a Maxwell field in the most general Einstein-Hilbert density Lagrangean with quadratic curvature invariants, to arrive at following equation
 \bq
\label{eq1} {\cal L}=\gamma R-\alpha
C_{\mu\nu\rho\sigma}C^{\mu\nu\rho\sigma}+\beta R^2-\kappa
F_{\mu\nu}F^{\mu\nu},
 \eq
where $F_{\mu\nu}=\nabla_\mu A_\nu-\nabla_\nu A_\mu$ is the electromagnetic tensor, $C_{\mu\nu\rho\sigma}$ is the Weyl tensor and $\alpha$,
$\beta$, $\gamma$ and $\kappa$ are coupling constants.

In the case analyzed in \cite{LPPS,WN}, the authors argued that the Ricci scalar should vanish, so that the equation of motion should not include the contribution from $\beta R^2$ term. The main argument behind  that conclusion is that the resulting tensor in the field equations that comes from the Weyl tensor is traceless. If we remember that Maxwell energy-momentum tensor is also traceless we can use the same arguments to conclude  that a charged black hole solution in this theory should not need of the contribution from $\beta R^2$ term. Thus, from now on we will set $\beta=0$ simplifying significantly the Einstein and Maxwell field equations that are reduced to
 \bqn
\label{eq2a}
R_{\mu\nu}-\frac{1}{2}g_{\mu\nu}R-4\alpha B_{\mu\nu}-2\kappa T_{\mu\nu}&=&0,\\
\nb\\
\label{eq2b}
\nabla_{\mu}F^{\mu\nu}&=&0.
\eqn
where $B_{\mu\nu}$ is the traceless Bach tensor and $T_{\mu\nu}$ is the energy-momentum tensor for the electromagnetic field,
given respectively by
\bqn
\label{eq3a}
B_{\mu\nu}&=&\left(\nabla^\rho\nabla^\sigma+\frac{1}{2}R^{\rho\sigma}\right)C_{\mu\rho\nu\sigma},\\
\nb\\
\label{eq3b}
T_{\mu\nu}&=&F_{\mu\alpha} F_{\nu}^{\phantom{\nu}\alpha} -\frac{1}{4}g_{\mu\nu}F_{\mu\nu}F^{\mu\nu}.
 \eqn

For sake of simplicity let us consider a general static spherically symmetric metric,
\bq
\label{eq4}
ds^2=-h(r)dt^2+\frac{dr^2}{f(r)}+r^2d\theta^2+r^2\sin^2(\theta)d\varphi^2.
 \eq
Substituting the Eq.(\ref{eq4}) into the field equations (\ref{eq2a}) and (\ref{eq2b}) one get the field equations,
 \bqn
\label{eq5}
 r h \left[r f' h'+2 f\left(r h''+2h'\right)\right]+4h^2 \left(r f'+f-1\right)-r^2 f h'^2&=&0,
\\
 \label{eq5a}
 f''+\frac{r^2 fh'^2+2 r f h h'+4 (f-1)h^2}{2 r f h \left(r h'-2 h\right)}f'-\frac{3 h f'^2}{4 f h-2 r f
 h'}& &\nb\\
 \nb\\
 +\frac{r^3f h'+\left(r^2f-r^2+\kappa Q_0^2\right)h}{\alpha r^2 f \left(r h'-2 h\right)} +\frac{r^3fh'^3-3r^2fhh'^2-8(f-1)h^3}{2r^2 h^2 \left(r h'-2
 h\right)}&=&0
 \eqn
 and
 \bqn
 \label{eq5b}
 A_t'+\sqrt{\frac{h}{f}}\frac{Q_0}{r^2}&=&0,
 \eqn
where $Q_0$ is interpreted as the electric charge since we are interested in asymptotically flat solutions. This requirement implies that the function $A_t$ has to vanish at infinity. If $Q_0=0$, these coupled equations reduce to those found in \cite{LPPS} and we recover the same solutions: the Schwarzschild and non-Schwarzschild black holes. On the other hand, despite the fact that the Schwarzschild metric be a solution of the system, when the Maxwell field was turned on, we found that the Reissner-Nordstr\"om metric was not a solution if $\alpha\not=0$. However, we will show that the system admits a new spherically symmetric  charged black hole solution.

\subsection{Asymptotic behaviour: the weak field limit}
\label{WFL}

Firstly, in order to gain some insights about the asymptotic behaviour of the charged black hole far from the region near to the event horizon we will analyze the Eqs.(\ref{eq5})-(\ref{eq5b}) in the weak field limit. Let us consider a linear approximation for $f(r),h(r)$ and $A_t(r)$, valid for sufficiently large $r$,  such that,
\bqn
\label{eq7a}
f(r)&=&1+f_\delta(r)+{\cal O}(f_\delta^2)\\
\nb\\
\label{eq7b}
h(r)&=&h_C(1+h_\delta(r))+{\cal O}(h_\delta^2),\\
\nb\\
\label{eq7c}
A_t(r)&=&A_\delta(r)+{\cal O}(A_\delta^2),
\eqn
and $f_\delta\sim h_\delta\sim A_\delta\ll 1$. Substituting the set of Eqs.(\ref{eq7a})-(\ref{eq7c}) into Eqs.(\ref{eq5})-(\ref{eq5b}),
the field equations are reduced to

\bqn
 \label{eq8a}
 \kappa Q_0^2+r^2f_\delta-4\alpha f_\delta+r^3 h_\delta'+2\alpha
 r^2f_\delta''&=&0,\\
\label{eq8b}
 f_\delta+rf_\delta'+rh_\delta'+\frac{1}{2}r^2h_\delta''&=&0, \\
\label{eq8c}
 A_\delta'+\sqrt{\frac{h}{f}}\frac{Q_0}{r^2}&=&0.
 \eqn

Solving the set of Eqs.(\ref{eq8a})-(\ref{eq8c}) we found the following asymptotic solutions for $f_\delta(r)$, $h_\delta(r)$ and $A_\delta(r)$

 \bqn
 \label{eq9}
 f_\delta(r)&=&-\frac{C_0}{r}-\left(r+\sqrt{2\alpha}\right)\frac{{\cal C}_1-\kappa Q_0^2E_i\left(\frac{r}{\sqrt{2\alpha}}\right)}{8\alpha r}e^{-\frac{r}{\sqrt{2\alpha}}}-\left(r-\sqrt{2\alpha}\right)\frac{{\cal C}_2-\kappa Q_0^2E_i\left(-\frac{r}{\sqrt{2\alpha}}\right)}{8\alpha r}e^{\frac{r}{\sqrt{2\alpha}}},\\
 \nb\\
 h_\delta(r)&=&-\frac{C_0}{r}-\frac{{\cal C}_1-\kappa Q_0^2E_i\left(\frac{r}{\sqrt{2\alpha}}\right)}{2\sqrt{2\alpha}r}e^{-\frac{r}{\sqrt{2\alpha}}}+\frac{{\cal C}_2-\kappa Q_0^2E_i\left(-\frac{r}{\sqrt{2\alpha}}\right)}{2\sqrt{2\alpha}r}e^{\frac{r}{\sqrt{2\alpha}}},\\
 \nb\\
 A_\delta(r)&=&\sqrt{h_C}\frac{Q_0}{r}+A_{\infty},
 \eqn
where $E_i(y)$ is the exponential integral function and $A_{\infty}$ is a constant that will be set to zero since we are interested in asymptotically flat solutions with electric potential vanishing at infinity. The constant $C_0$ in the coefficient of $1/r$ term can be interpreted as the black hole's mass $M$ if we set $\mathcal{C}_1 = Q_0 =0$. About the two constants ${\cal C}_1$ and ${\cal C}_2$ one can see that, at infinity, the terms related to ${\cal C}_2$ are exponentially divergent while the terms related to $C_1$ goes exponentially to zero. Thus we must set ${\cal C}_2=0$. To understand what the role played by $C_1$ we will put these solutions in a more illustrative format further expanding for large $r$. They become

\bqn
 \label{eq10}
 f(r)&=&1-\frac{2M-\frac{\kappa Q_{0}^2}{2 \sqrt{2\alpha}}}{r}+\frac{3\sqrt{2\alpha}\kappa Q_{0}^2}{2 r^3}-e^{-\frac{r}{\sqrt{2\alpha}}}\ {\cal C}_1 \left(\frac{1}{8\alpha}+\frac{1}{4\sqrt{2\alpha} r}\right)+\mathcal{O}\left(1/r^5\right),\\
 \nb\\
 \label{eq10a}
 h(r)&=&h_C \left[1-\frac{2M}{r}+\frac{\kappa Q_{0}^2 }{r^2}+\frac{4 \alpha k Q_{0}^2 }{r^4}-e^{-\frac{r}{\sqrt{2\alpha}}}\left(\frac{{\cal C}_1}{2\sqrt{2\alpha} r}\right)\right] +\mathcal{O}\left(1/r^5\right),\\
 \nb\\
 A_t(r)&=&\sqrt{h_C}\frac{Q_0}{r}.
 \eqn

An inspection in the Eq.(\ref{eq10}) reveals that the effective mass $\mathcal{M}=M-\frac{\kappa Q_{0}^2}{2 \sqrt{2\alpha}}$ can assume negative values if the electric charge is large enough. This behaviour was already pointed out in \cite{LPPS} for non-Schwarzschild black holes when $r_0>r_{0}^{m=0}=1.143$ where $r_0^{m=0}$ is a critical value for the event horizon. On the other hand, the Eq.(\ref{eq10a}) has a behaviour quite similar that of the Reissner-Nordstr\"om metric. In both functions the term with $\mathcal{C}_1$ is exponentially damped when $r\to\infty$ indicating that this ``geometric hair" of the black hole is almost undetectable far from the event horizon region. The weak field limit for higher derivative curvature non-charged black hole solutions was discussed in \cite{chinese} and our analysis recover their results when $Q_0$ is vanishing.
\begin{figure}[t]
\includegraphics[width=5.5cm]{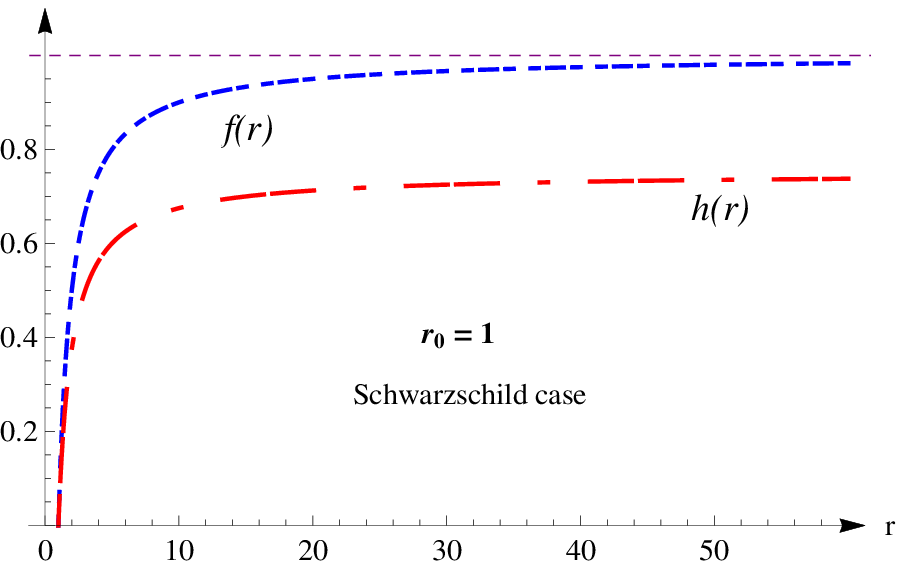}\includegraphics[width=5.5cm]{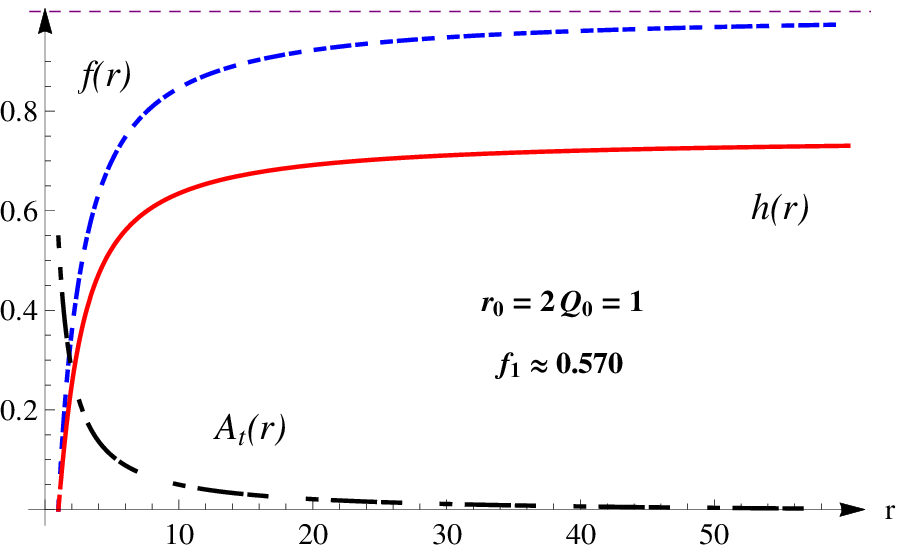}\includegraphics[width=5.5cm]{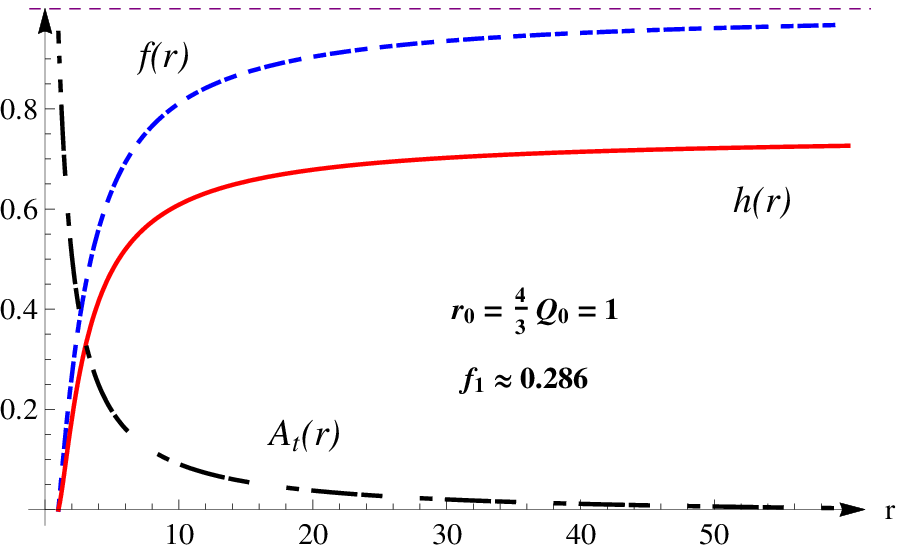}
\includegraphics[width=5.5cm]{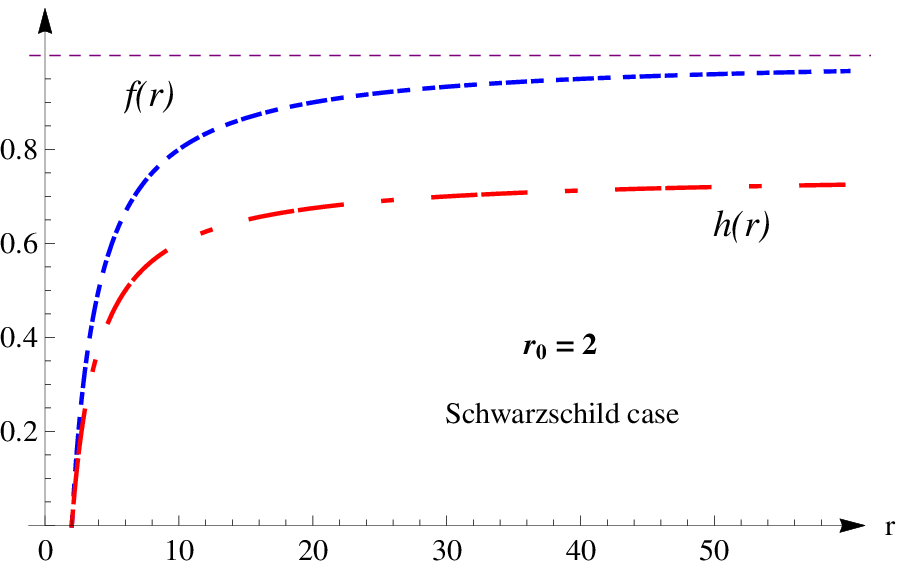}\includegraphics[width=5.5cm]{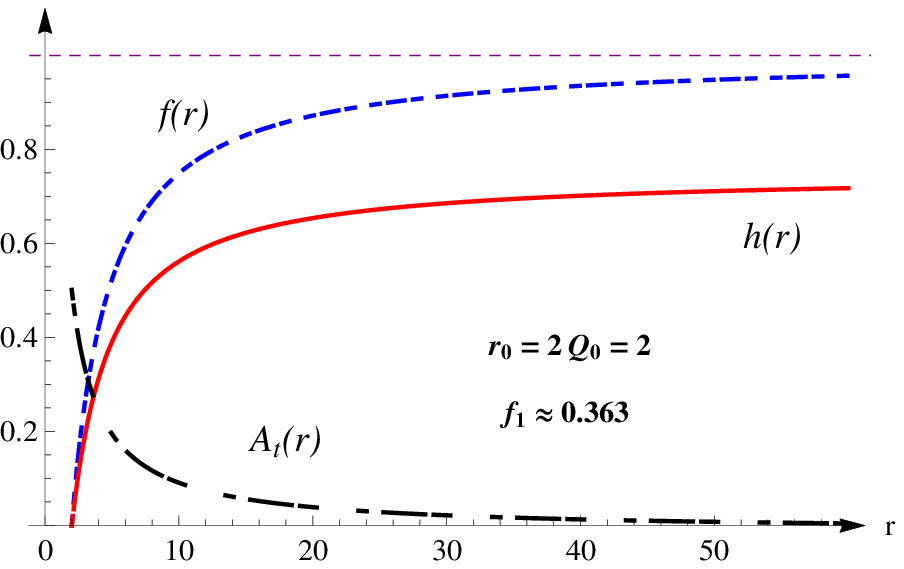}\includegraphics[width=5.5cm]{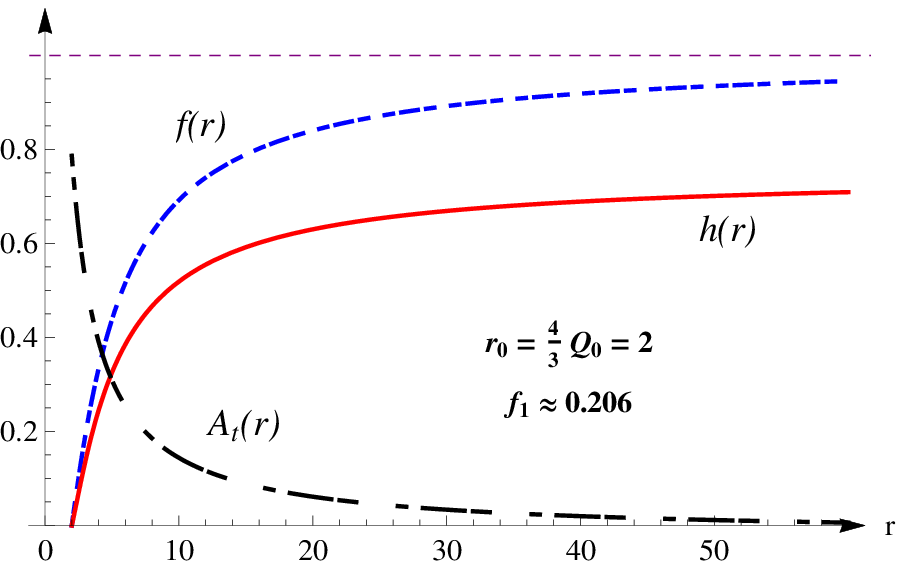}
\includegraphics[width=5.5cm]{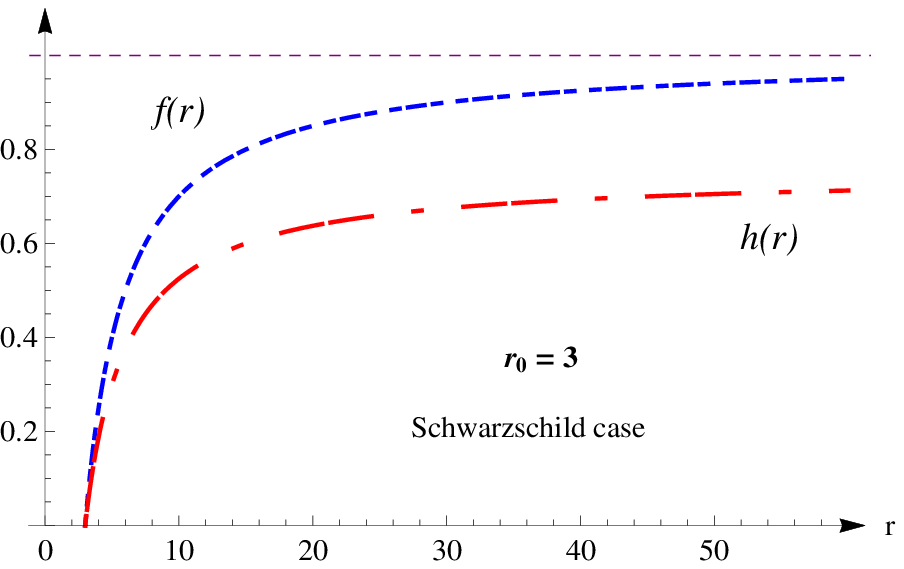}\includegraphics[width=5.5cm]{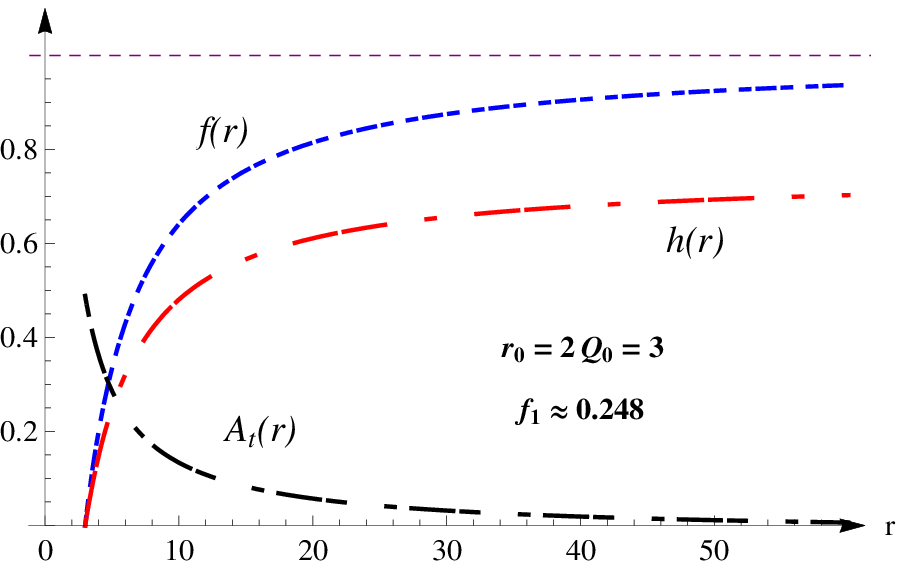}\includegraphics[width=5.5cm]{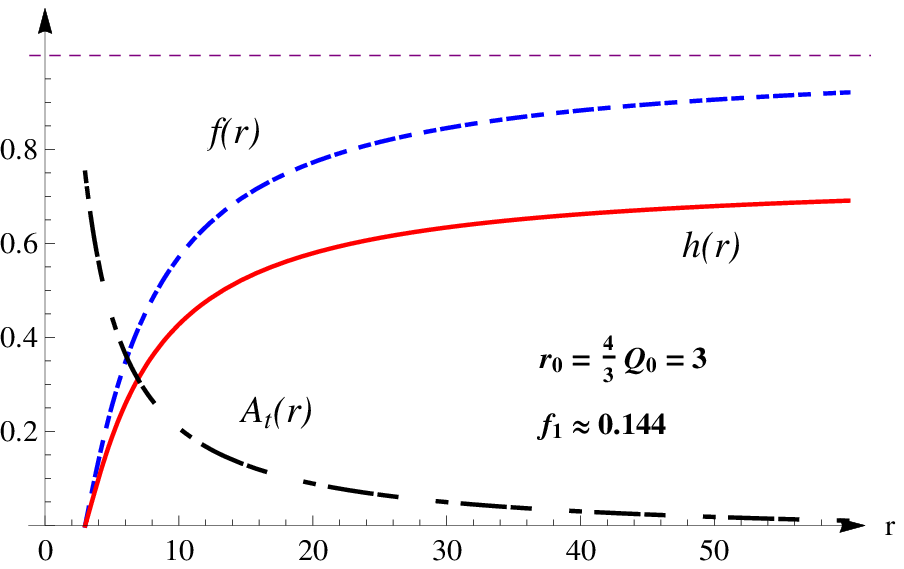}
\caption{Numerical solutions of the Group I for $f(r)$, $h(r)$ and $A_t(r)$ with $r_0=1,2,3,4/3$ and some values of $Q_0$. In each plot the function $h$ was chosen to approach to $3/4$ instead of 1 for clarity. The purple dashed line represents the unity.} \label{fig2a}
\end{figure}

\subsection{Numerical black hole solutions}
In general, charged black holes have more than one horizon because of the functional form of influence of the electric charge in the metric. For example, the Reissner-Nordstr\"om black hole has one event horizon and one Cauchy horizon. As to find analytic solutions from Eqs.(\ref{eq5}-\ref{eq5b}) it is not an easy task we decide to compute numerical solutions. Here we will suppose that the spacetime has only one horizon to make easier the expansion of $f(r)$ and $h(r)$ around the event horizon $r_0$. Thus, $h(r)$ and $f(r)$ become
\bqn
\label{eq6}
 h(r)&=&h_1(r-r_0)+h_2(r-r_0)^2+h_3(r-r_0)^3+\cdot\cdot\cdot,\\
 f(r)&=&f_1(r-r_0)+f_2(r-r_0)^2+f_3(r-r_0)^3+\cdot\cdot\cdot,
\label{eq6a}
 \eqn
where $f_i$ and $h_i$ are constant coefficients near to the event horizon. Moreover, since one can always rescale the time coordinate, we set $h_1=f_1$ for the sake of convenience in following calculations. Substituting the expansions (\ref{eq6})-(\ref{eq6a}) into Eqs.(\ref{eq5})-(\ref{eq5b}), all $h_j$ and $f_j$ with $j \ge 2$ can be calculated from $f_1$. For example, $h_2$ and $f_2$ can be written as
 \bqn
\label{eq7}
 h_2&=&\frac{1-2f_1r_0}{r_0^2}-\frac{r_0^2-f_1r_0^3-\kappa Q_0^2}{8\alpha f_1r_0^3},\nonumber\\
 \\
 f_2&=&\frac{1-2f_1r_0}{r_0^2}-3\frac{r_0^2-f_1r_0^3-\kappa Q_0^2}{8\alpha f_1r_0^3}.\nonumber
 \eqn

The black hole solutions will depend of three free parameters, so that the event horizon $r_0$, electric charge $Q_0$ and $f_1$.
The parameter $\kappa$ will be fixed in the next section when we present the numerical black hole solutions.
The integration of the equations of motion were performed using numerical routines in MATHEMATICA and it was executed in the range $r_0 < r <r_L$ where $r_L$ is a sufficiently large value of $r$. In the expansions for $h(r)$ and $f(r)$ the terms of
order $\mathcal{O}[(r-r_0)^9]$ were discarded.

\begin{figure}[t]
\includegraphics[width=5.5cm]{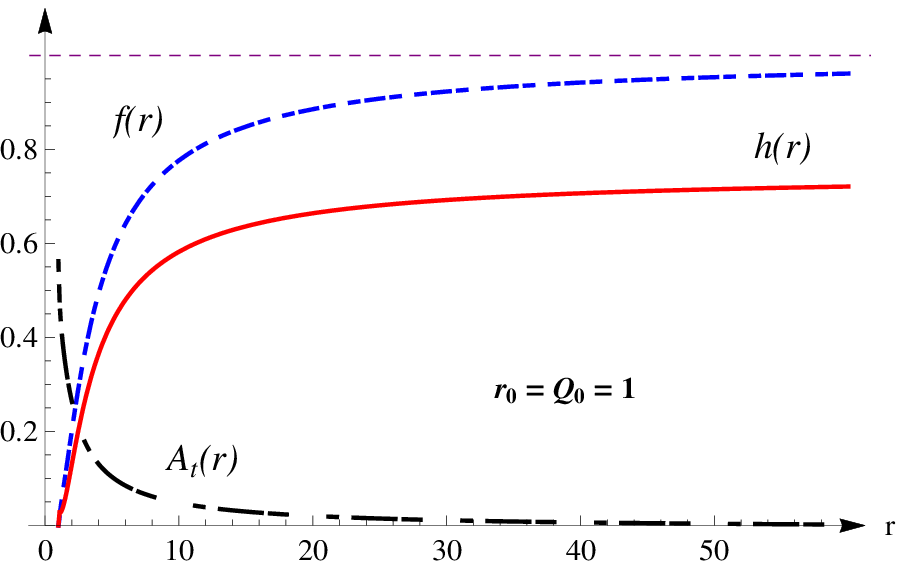}\includegraphics[width=5.5cm]{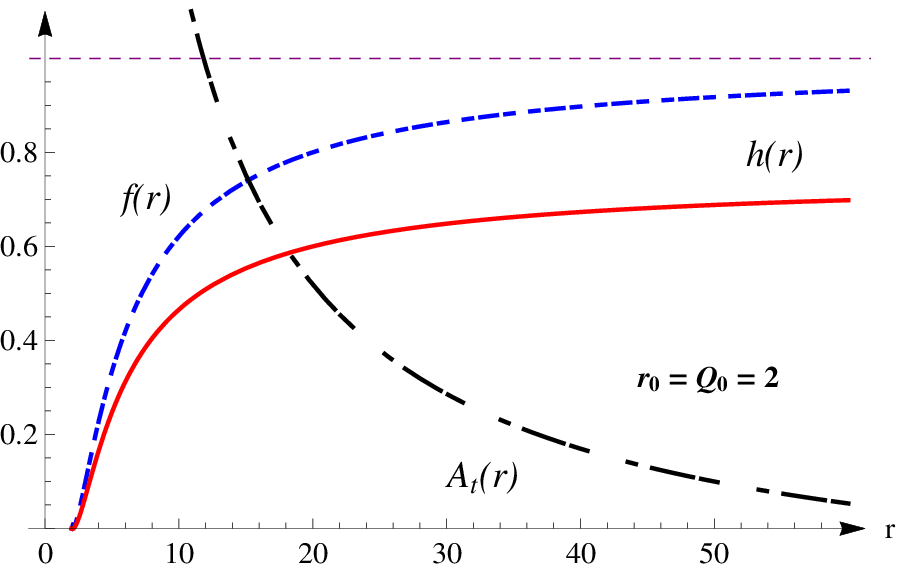}\includegraphics[width=5.5cm]{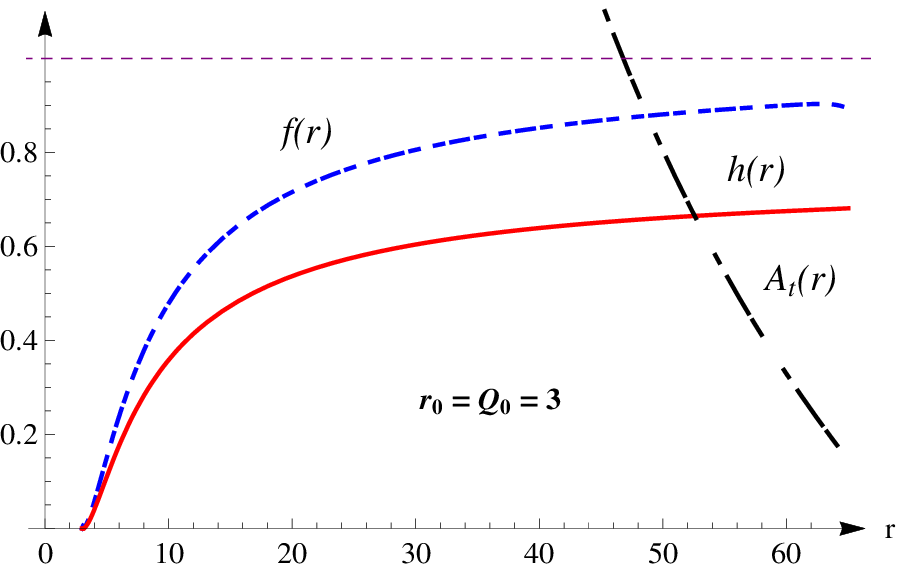}
\caption{Numerical solutions of the group I for $f(r)$, $h(r)$ and
$A_t(r)$ for some values of $r_0$ - (extreme cases). In each plot the function $h$ was chosen to
approach to $3/4$ for clarity. The purple dashed line represents the unity.} \label{fig2b}
\end{figure}

From now on we shall take $\alpha=\frac12$ and $\kappa=1$ without loss of generality. So, choosing suitable values for $f_1$, $r_0$ and $Q_0$ and integrating the equations of motion with \emph{NDSolve} routine we were able to find two groups of charged black hole solutions which have a large enough region outside of the event horizon.

The charged black hole solutions will be separated in two groups, according to the non-charged ``seed" solution. The first one,
named {\bf Group I}, could be viewed as a charged generalization of the higher derivative curvature Schwarzschild black hole considered in \cite{LPPS}.  Although the Schwarzschild metric is solution of the field equations if $f_1=1$, the Reissner-Nordstr\"om metric will not be. In Fig.(\ref{fig2a}) we plot some solutions of Group I for some values $r_0$, $Q_0$ and $f_1$.

\begin{figure}[h]
\includegraphics[width=5.5cm]{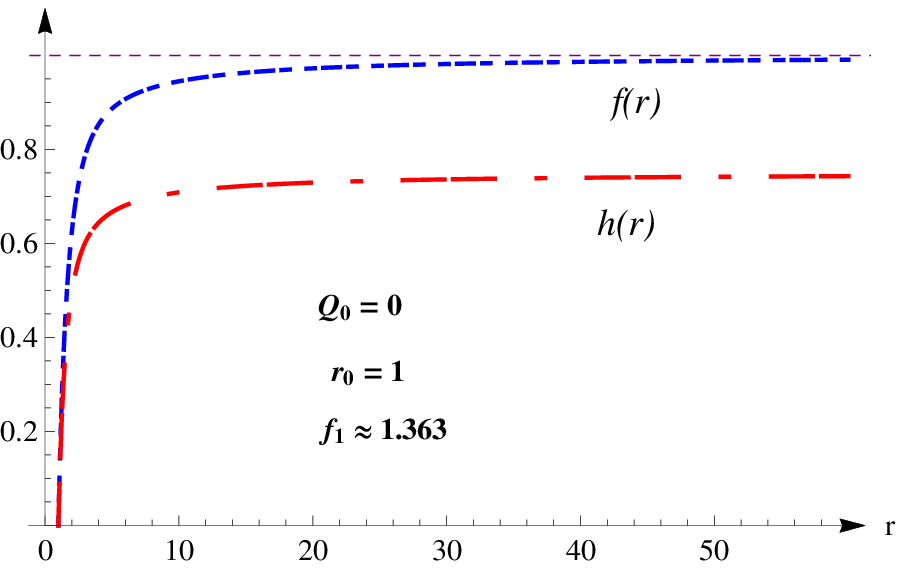}\includegraphics[width=5.5cm]{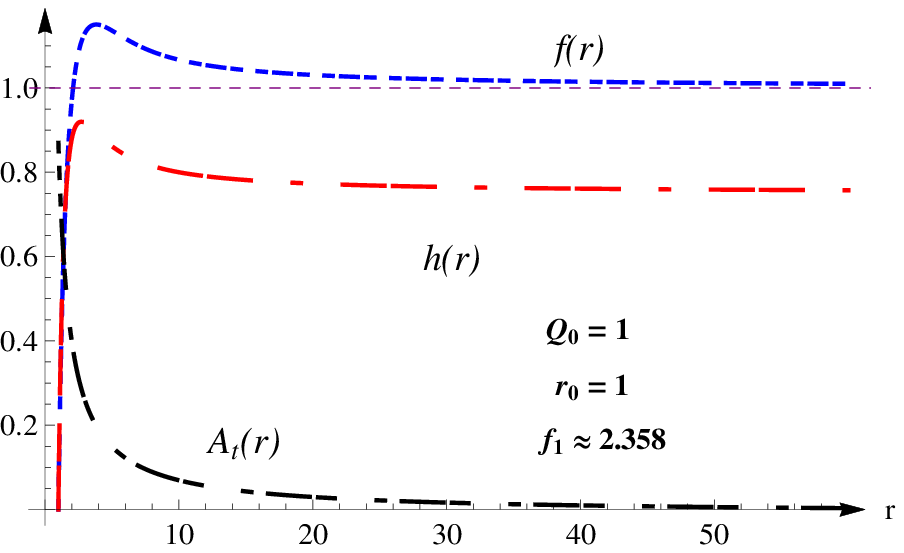}\includegraphics[width=5.5cm]{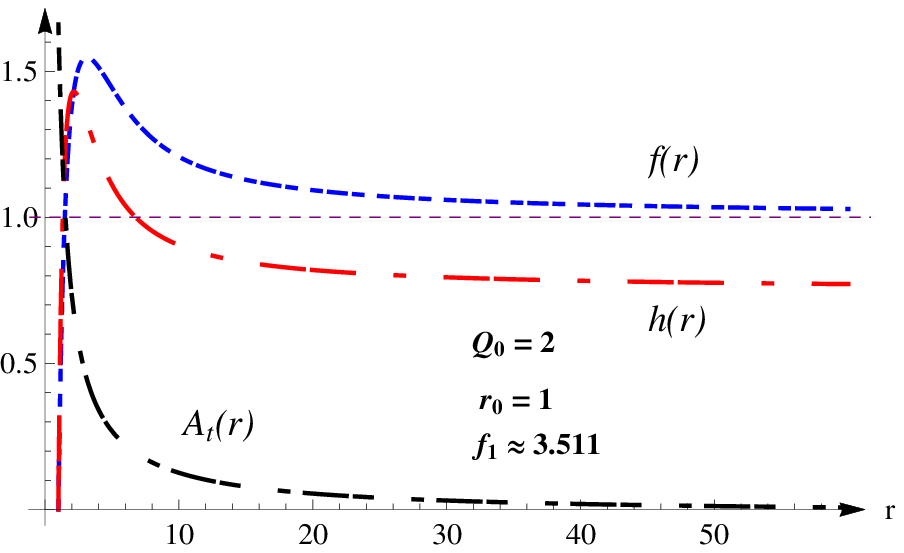}
\includegraphics[width=5.5cm]{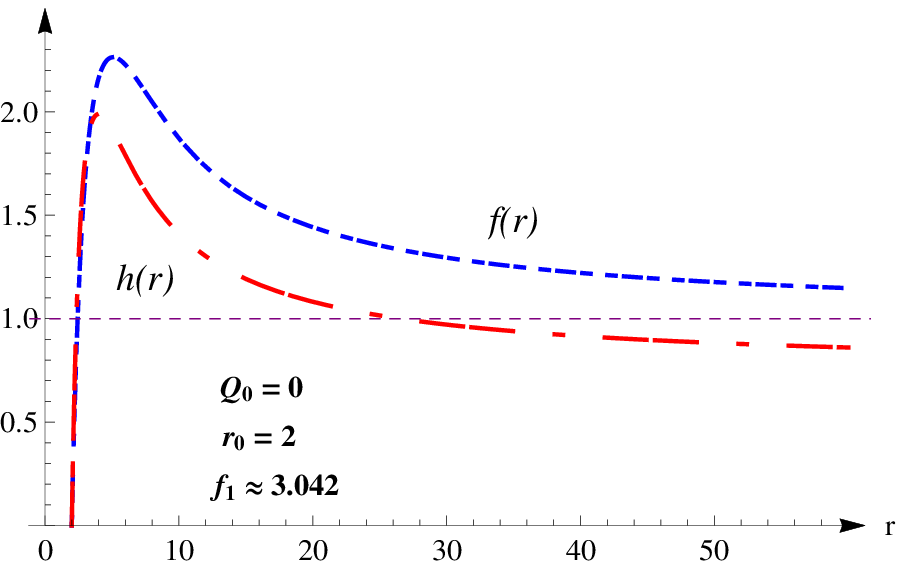}\includegraphics[width=5.5cm]{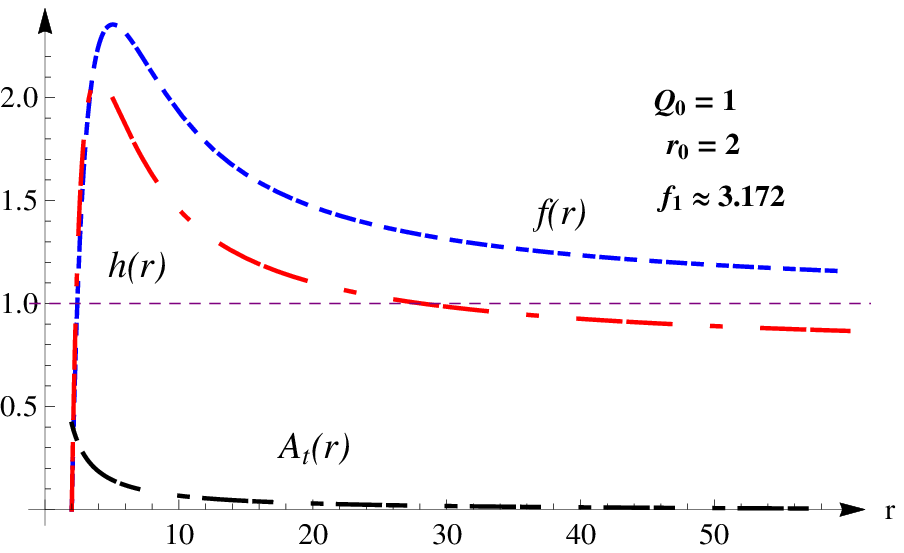}\includegraphics[width=5.5cm]{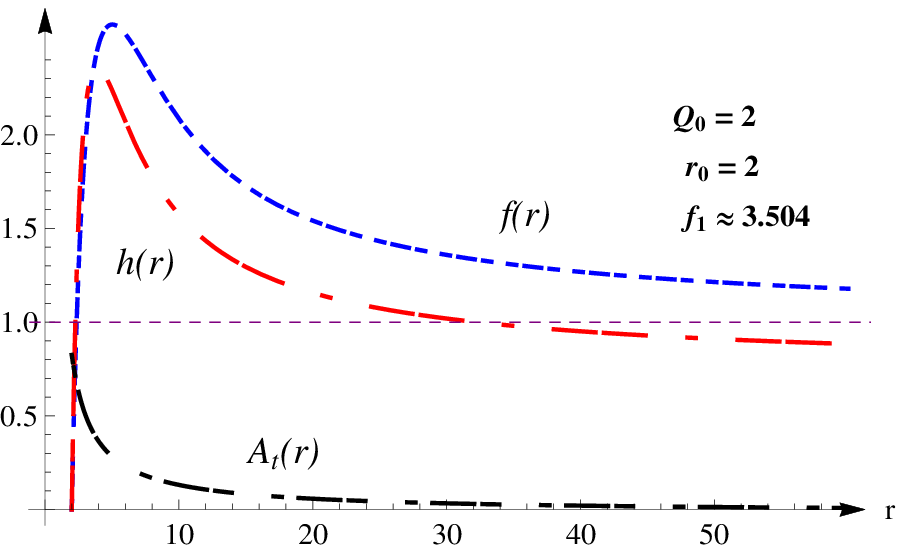}
\includegraphics[width=5.5cm]{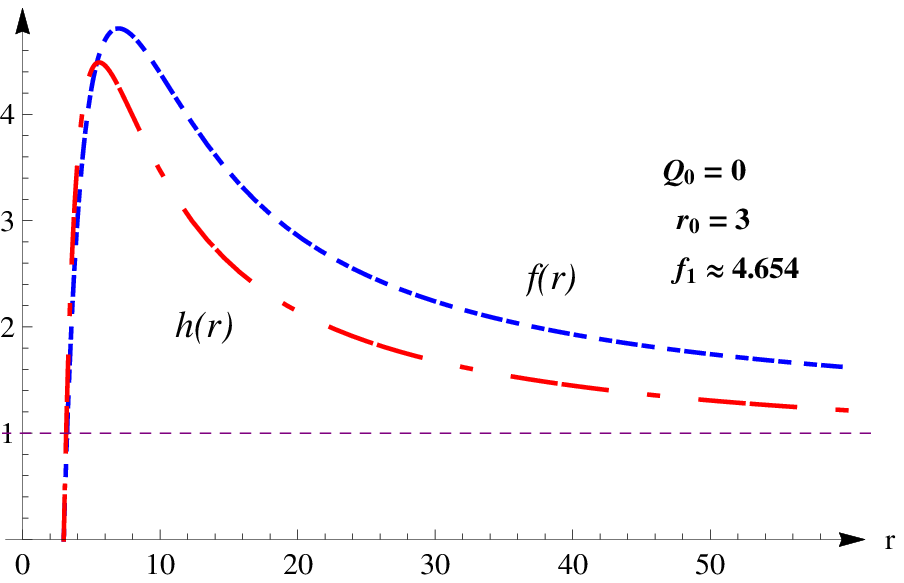}\includegraphics[width=5.5cm]{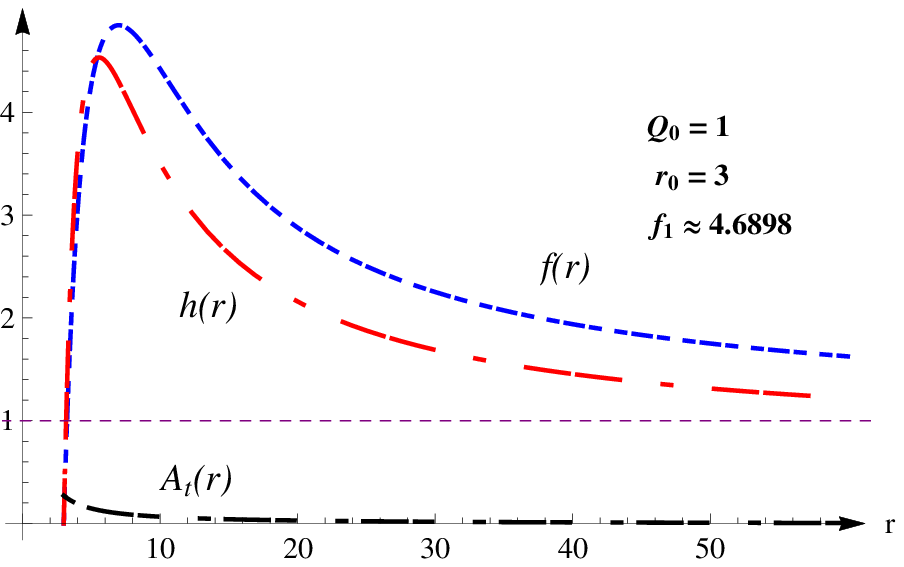}\includegraphics[width=5.5cm]{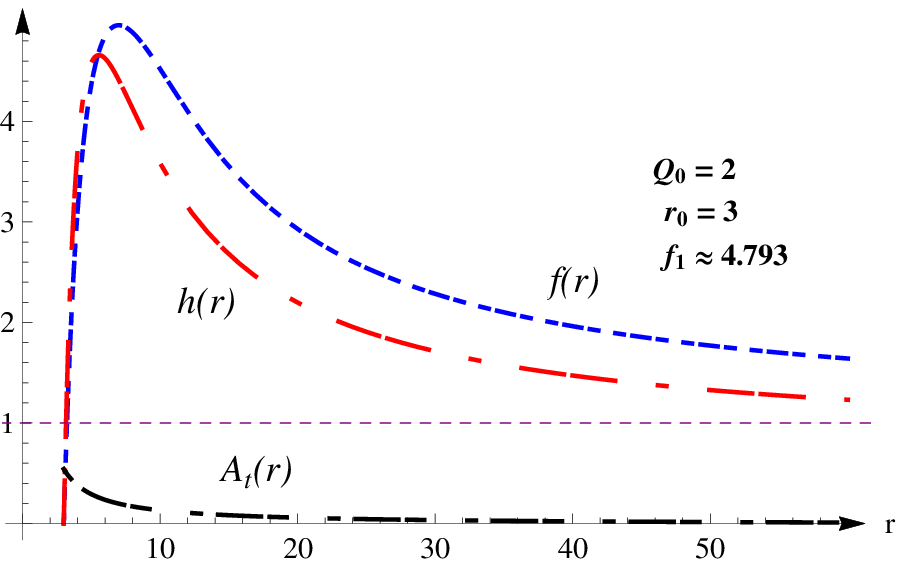}
\caption{Numerical solutions of group II for $f(r)$, $h(r)$ and $A_t(r)$ with $r_0=1,2,3$ and $Q_0=0,1,2$. In each plot the function $h$ was chosen to approach $3/4$ for clarity. The purple dashed line represents the unity.} \label{fig1c}
\end{figure}

The behaviour of these black hole solutions is quite similar that Schwarzschild metric one. The functions $f$ and $h$ are increasing functions in the region $r_0\leq r < \infty$. The potential vector $A_t$ is finite at the horizon and decays to zero far from the black hole. Extreme charged black hole solutions could exist when we set $r_0 =Q_0$ and $f_1=0$. Under these conditions one can see in the Eqs.(\ref{eq7})that the black holes has a vanishing temperature as expected in this case. Some examples of extreme charged black hole solutions are depicted in Fig.(\ref{fig2b}).

The solutions of {\bf Group II} could be viewed as a higher derivative curvature electrically charged generalization of non-Schwarzschild solution. In the functions $f$ and $h$ of these solutions appear a peak outside the event horizon that can be related to the presence of a negative effective mass just like discussed in \cite{LPPS} for the non-Schwarzschild solution. There are not extreme charged black hole solutions for this group. We draw the $f$, $h$ and $A_t$ for the black hole metrics with $r_0=1,2,3$ and $Q_0=0,1,2$ in Fig.(\ref{fig1c}).

Differently of the non-Schwarzschild solution discussed in \cite{LPPS} where it was possible to find the critical value for the event horizon $r_0^{m=0}=1.143$ that separates the two groups of solutions, here, it was not. Probably the reason for this difficulty is because we have to set three parameters to define our black hole solutions.

\begin{figure}[h]
\includegraphics[width=6cm]{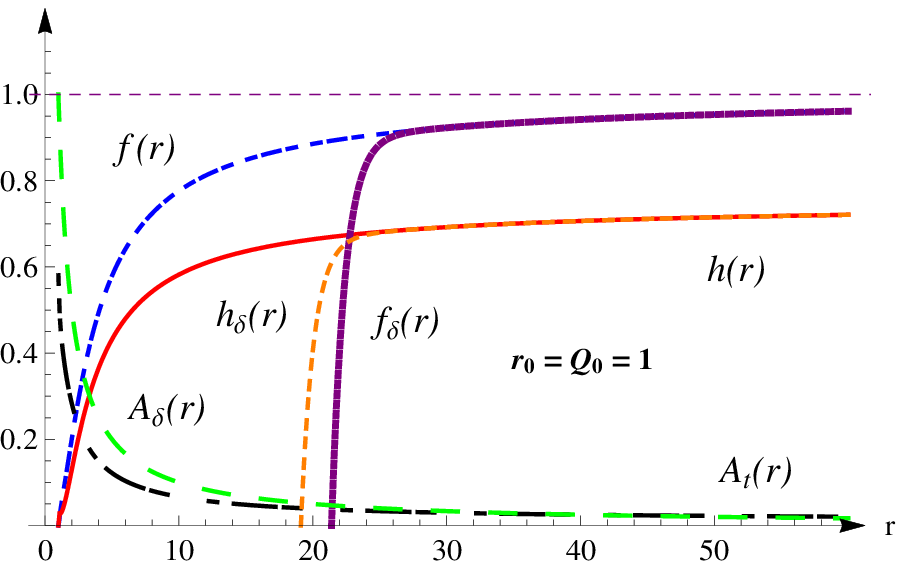}\includegraphics[width=6cm]{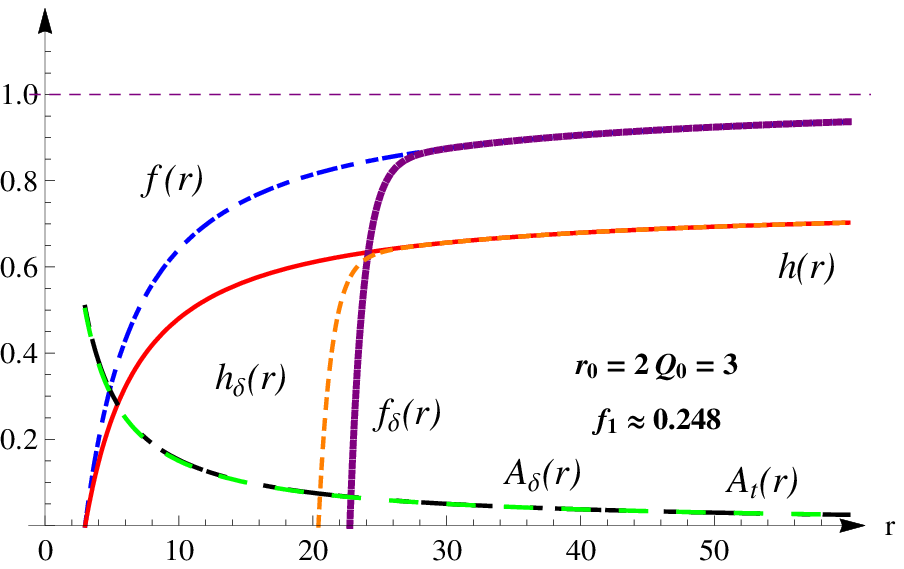}\includegraphics[width=6cm]{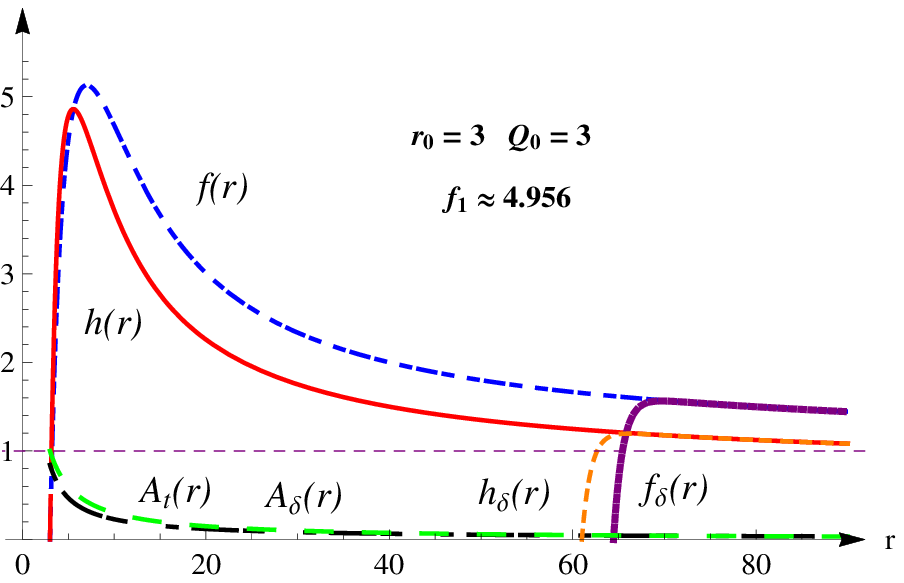}
\caption{Numerical solutions and the weak field limit for $f(r)$, $h(r)$ and $A_t(r)$ for some values of $r_0$. The purple dashed line represents the unity. The purple, orange and green line represent the weak field limit functions derived in the section \ref{WFL}} \label{fig2W}
\end{figure}

In order to compare the asymptotic behaviour of our numerical solutions and the weak field limit equations (\ref{eq10},\ref{eq10a}) we plot the Fig(\ref{fig2W}). As one can see they are in good agreement.

\section{Conclusion}
\label{conclusions}

In this paper, we obtained numerically, electrically charged asymptotically flat black hole solutions in Einstein-Hilbert-Maxwell with quadratic curvature invariants gravity. These solutions were separated in two groups according to their seed solutions: Schwarzschild and non-Schwarzschild. Extreme black hole solutions and black holes with negative mass were found. Their proprieties and how to obtain these black hole solutions were discussed in detail.

We expect to address in future work the possibility of using black holes of higher derivative gravity as
particle accelerator, as proposed by Ba\~nados, Silk and West in \cite{BSW}, where they proved that the ``extreme Kerr black hole accelerator" can reach arbitrarily high energy level in principle for suitable conditions. That interesting work immediately attracted the attention of many physicists resulting in several related works in other scenarios \cite{Harada,Jacobson, Zaslavskii,Berti, Kimura,Wei,Yao,Wei1,Zhu,Zaslavskii1,Abduj,Nakao, Patil,Patil1,Harada1,Amir,Pradhan,OBZ,OBZ2}. Following this previous idea one could ask if the same mechanism appears in this higher derivative theory of gravity.

Another interesting possibility is the alternative theory of gravity of Ho\v{r}ava-Lifshitz. Recently, Ho\v{r}ava proposed a new idea for a theory of gravity where the Lorentz symmetry could be broken at high energy region \cite{HL}.
According to this idea, Ho\v{r}ava-Lifshitz gravity and several
modified Ho\v{r}ava-Lifshitz gravity were proposed, and it is proved
that ghost problem, instability and strong coupling difficulty can
be canceled in the new renormalizable gravity \cite{HLmais}. Besides, the post-newtonian approximation also won't exclude this
theory. Therefore, would be very interesting to investigate the charged and rotating black holes as particle accelerators in
Ho\v{r}ava-Lifshitz gravity \cite{PostNewtonian}.

Recently, black hole solutions in Lorentz symmetry breaking theories have received much attention \cite{HL,HLmais,Lorentzviolation} in special because, despite the fact that the concept of event horizon seems to be meaningless in these theories, they have a similar concept named ``universal horizon" whose have the propriety that to trap any particle even with arbitrarily high speed\cite{UH}. Fortunately, the universal horizon lies inside the event horizon, thus a particle interaction at very high energy can produce an outcome that escapes of the black hole. Additionally, in modified gravity without Lorentz symmetry, the Hamilton-Jacobi equation as well as the geodesic equation get, both, modified too. Therefore, we have to reexamine their solutions in the case of the black hole accelerator. This subject is currently under study.

\section*{\bf Acknowledgements}

This work is supported in part by FAPESP No. 2012/08934-0, Conselho Nacional de Desenvolvimento Cient\'{\i}fico e Tecnológico (CNPq-Brazil), grant 472660/2013-6, CAPES, and NNSFC No.11573022 and No.11375279.

\onecolumngrid


\begin{thebibliography}{nbound}

\bibitem{Stelle} K.S.  Stelle, Phys. Rev. D {\bf 16}, 953 (1977).

\bibitem{Pais} A. Pais and G.E. Uhlenbeck, Phys. Rev. {\bf 79}, 145(1950).

\bibitem{Smilga} A.V.  Smilga, J. Phys. A{\bf 47}, (2014) 052001, [arXiv:1306.6066 ]

\bibitem{Smilga1} A. V.  Smil  Nucl. Phys. B{\bf 706}, 598 (2005). [hep-th/0407231]

\bibitem{LPPS}  H. L\"u, A.Perkins, C.N.Pope and K.S.Stelle, Phys.Rev.Lett. {\bf 114}, 171601 (2015) [arXiv:1502.01028 [hep-th]];
Phys. Rev. D {\bf 92}, 124019 (2015) [arXiv:1508.00010 [hep-th]].


\bibitem{WN}  W. Nelson, Phys.Rev.D {\bf 82}, 104026 (2010).

\bibitem{chinese} Y.~F.~Cai, G.~Cheng, J.~Liu, M.~Wang and H.~Zhang, JHEP {\bf 1601}, 108 (2016) [arXiv:1508.04776 [hep-th]].





\bibitem{BSW}  M. Ba\~nados, J.Silk and S.M. West, Phys.Rev.Lett. {\bf 103}, 111102 (2009); M. Ba\~nados, B. Hassanain, J. Silk, and S. M. West, Phys. Rev. D {\bf 83}, 023004 (2011) [arXiv:1010.2724 [astro-ph]].

\bibitem{Harada} T. Harada and M. Kimura, Phys. Rev. D {\bf 83}, 024002 (2011) [arXiv:1010.0962 [gr-qc]];
Phys. Rev. D {\bf 83}  084041 (2011) [arXiv:1102.3316 [gr-qc]].

\bibitem{Jacobson} T. Jacobson and T. P. Sotiriou, Phys. Rev. Lett. {\bf 104},  021101 (2010) [arXiv:0911.3363 [gr-qc]].

\bibitem{Zaslavskii} O. B. Zaslavskii, Phys. Rev. D {\bf 84}, 024007 (2011) [arXiv:1104.4802 [gr-qc]].

\bibitem{Berti} E. Berti, V. Cardoso, L. Gualtieri, F. Pretorius, and U. Sperhake, Phys. Rev. Lett. {\bf 103}, 239001 (2009)
[arXiv:0911.2243 [gr-qc]].

\bibitem{Kimura} M. Kimura, K. -i. Nakao, and H. Tagoshi, Phys. Rev. D {\bf 83}, 044013 (2011) [arXiv:1010.5438 [gr-qc]].

\bibitem{Wei} S. W. Wei, Y. X. Liu, H. Guo, and C. E. Fu, Phys. Rev. D {\bf 82}, 103005 (2010) [arXiv:1006.1056 [hep-th]].

\bibitem{Yao} W. Yao, S. Chen, C. Liu, and J. Jing, Eur. Phys. J. C {\bf 72}, 1898 (2012) [arXiv:1105.6156 [gr-qc]].

\bibitem{Wei1} S. W. Wei, Y. X. Liu, H. T. Li, and F. W. Chen, JHEP {\bf 1012}, 066 (2010) [arXiv:1007.4333 [hep-th]].

\bibitem{Zhu} Y. Zhu, S. -F. Wu, Y. -X. Liu, and Y. Jiang, Phys. Rev. D {\bf 84}, 043006 (2011) [arXiv:1103.3848 [hep-th]].

\bibitem{Zaslavskii1} O. B. Zaslavskii, Phys. Rev. D {\bf 82},  083004 (2010).

\bibitem{Abduj} A. Abdujabbarov, N. Dadhich, B. Ahmedov and H. Eshkuvatov, Phys. Rev. D {\bf 88}, 084036 (2013)
 [arXiv:1310.4494[gr-qc]].

\bibitem{Nakao} K. -i. Nakao, M. Kimura, M. Patil and P. S. Joshi, Phys. Rev. D {\bf 87},  104033 (2013) [arXiv:1301.4618 [gr-qc]].

\bibitem{Patil} M. Patil and P. S. Joshi, Class. Quant. Grav. {\bf 28},  235012 (2011) [arXiv:1103.1082 [gr-qc]].

\bibitem{Patil1} M. Patil and P. S. Joshi, Phys. Rev. D {\bf 84},  104001 (2011) [arXiv:1103.1083 [gr-qc]].

\bibitem{Harada1} T. Harada and M. Kimura, Class. Quan. Grav. {\bf 31}, 243001 (2014).

\bibitem{Amir} M. Amir and S. G. Ghosh, JHEP {\bf 1507}, 015 (2015).

\bibitem{Pradhan} P. Pradhan, {\it Regular Black Holes as Particle Accelerators}, arXiv:1402.2748v3[gr-qc].

\bibitem{OBZ}  O.B.Zaslavskii, JETP Lett. {\bf 92}, 571 (2010).

\bibitem{OBZ2} O.B.Zaslavskii, {\it Schwarzschild black hole as particle accelerator of spinning particles},  arXiv:1603.09353 [gr-qc].

\bibitem{HL} P. Ho\v{r}ava, Phys. Rev. D {\bf 79},   084008 (2009) [arXiv:0901.3775 [hep-th]].

\bibitem{HLmais} A. M. da Silva, Class. Quantum Grav. {\bf 28}, 055011 (2011);
P. Ho\v{r}ava and C. M. Melby-Thompson, Phys. Rev. D {\bf 82}  064027 (2010);
A. Wang, Phys. Rev. D {\bf 82}, 124063 (2010), [arXiv:1008.3637 [hep-th]];
 A. Borzou, K. Lin and A. Wang, J. Cosmol. Astropart. Phys. {\bf 05}, 006 (2011) [arXiv:1103.4366[hep-th]];
 S. Mukohyama, J. Cosmol. Astropart. Phys. {\bf 06}, 001 (2009), [arXiv:0904.2190 [hep-th]];
T. Zhu, Q. Wu, A. Wang and F.-W. Shu, Phys. Rev. D {\bf 84},  101502 (2011) [arXiv:1108.1237 [hep-th]];
T. Zhu, F.-W. Shu, Q. Wu and A. Wang, Phys. Rev. D {\bf 85},  044053 (2012)[ arXiv:1110.5106 [hep-th]];
A. Wang and Y. Wu, Phys. Rev. D {\bf 83}, 044031 (2011)  [arXiv:1009.2089 [hep-th]];
K. Lin, A. Wang, Q. Wu and T. Zhu, Phys. Rev. D {\bf 84},  044051 (2011), [arXiv:1106.1486 [hep-th]];
E. Abdalla and A. M. da Silva, Phys. Lett. B {\bf 707}, 311 (2012).

\bibitem{PostNewtonian} K. Lin, S. Mukohyama and A.Wang, Phys. Rev. D {\bf 86}, 104024 (2012) [arXiv:1206.1338 [hep-th]];
K. Lin and A. Wang, Phys. Rev. D {\bf 87}, 084041 (2013) [arXiv:1212.6794 [hep-th]];
K. Lin, S. Mukohyama, A. Wang and T. Zhu, Phys. Rev. D 89, 084022 (2014) [arXiv:1310.6666 [hep-ph]].

\bibitem{Lorentzviolation} J. Magueijo and L. Smolin, Class.Quant.Grav. {\bf 21},  1725 (2004) [arXiv:gr-qc/0305055];
T. Jacobson, S. Liberati and D. Mattingly, Phys.Rev. D {\bf 67}, 124011 (2003) [arXiv:hep-ph/0209264]

\bibitem{UH} D. Blas and S. Sibiryakov, Phys. Rev. D {\bf 84}, 124043 (2011);
K. Lin, E. Abdalla, R.-G. Cai, and A. Wang, Int. J. Mod. Phys. D {\bf 23}, 1443004 (2014);
K. Lin, O. Goldoni, M. F. da Silva and A. Wang, Phys. Rev. D {\bf 91}, 024047 (2015);
K. Lin, F-W Shu, A. Wang and Q. Wu, Phys. Rev. D {\bf 91}, 044003 (2015).

\end{thebibliography}
\end{document}